\documentclass[11pt]{article}

\usepackage{graphics} % for pdf, bitmapped graphics files 
\usepackage{grffile}
\usepackage{subcaption}
\usepackage[margin=1.2in]{geometry}

\usepackage{epsfig} % for postscript graphics files
\usepackage{mathptmx} % assumes new font selection scheme installed
\usepackage{times} % assumes new font selection scheme installed
\usepackage{amsmath} % assumes amsmath package installed
\usepackage{amssymb,amsfonts}  % assumes amsmath package installed
  
\usepackage{hyperref}

\newcommand{\fdr}{\textnormal{FDR}}
\newcommand{\nulls}{\mathcal{H}^0}
\newcommand{\propref}[1]{Proposition~\ref{prop:#1}}
\newcommand{\thmref}[1]{Theorem~\ref{thm:#1}}
\newcommand{\secref}[1]{Section~\ref{sec:#1}}
\newcommand{\appref}[1]{Appendix~\ref{app:#1}}
\newcommand{\figref}[1]{Figure~\ref{fig:#1}}
\newcommand{\PP}[1]{\textnormal{Pr}\!\left\{{#1}\right\}} % Probability
\newcommand{\PPst}[2]{\textnormal{Pr}\!\left\{{#1}\ \middle| \ {#2}\right\}} % Conditional probability
\newcommand{\EE}[1]{\mathbb{E}\left[{#1}\right]} % Expectation
\newcommand{\V}{\mathcal{V}}
\newcommand{\E}{\mathcal{E}}
\newcommand{\HH}{\mathcal{H}}
\newcommand{\kh}{\widehat{k}}
\newcommand{\bh}{\textnormal{BH}}
\newcommand{\Dset}{\ensuremath{\mathcal{D}}}
\newcommand{\One}[1]{{\bf{1}}\left\{{#1}\right\}}
\newcommand{\prank}{\text{$p$-rank}}
\newcommand{\N}{\mathbb{N}}

\newtheorem{theorem}{Theorem}

\newtheorem{proposition}{Proposition}

\newtheorem{definition}{Definition}

\makeatletter

\newcommand{\newreptheorem}[2]
{\newenvironment{rep#1}[1]
{\def\rep@title{#2 \ref{##1}} \begin{rep@theorem}}%
 {\end{rep@theorem}}}
\makeatother
\newreptheorem{theorem}{Theorem}
\newreptheorem{lemma}{Lemma}
\newreptheorem{corollary}{Corollary}
\newreptheorem{proposition}{Proposition}

\makeatletter
\newcommand{\dotfrac}[2]{
\mathchoice
{\ooalign{$\genfrac{}{}{0pt}{0}{#1}{#2}$\cr\leavevmode\cleaders\hb@xt@ .22em{\hss $\displaystyle\cdot$\hss}\hfill\kern\z@\cr}}
{\ooalign{$\genfrac{}{}{0pt}{1}{#1}{#2}$\cr\leavevmode\cleaders\hb@xt@ .22em{\hss $\textstyle\cdot$\hss}\hfill\kern\z@\cr}}
{\ooalign{$\genfrac{}{}{0pt}{2}{#1}{#2}$\cr\leavevmode\cleaders\hb@xt@ .22em{\hss $\scriptstyle\cdot$\hss}\hfill\kern\z@\cr}}
{\ooalign{$\genfrac{}{}{0pt}{3}{#1}{#2}$\cr\leavevmode\cleaders\hb@xt@ .22em{\hss $\scriptscriptstyle\cdot$\hss}\hfill\kern\z@\cr}}
}
\makeatother

\title{\LARGE \bf
QuTE: decentralized multiple testing 
on sensor networks
 with false discovery rate control\footnote{This paper appeared in the IEEE CDC'17 conference proceedings \cite{ramdas2017qute}. The last two sections were then developed in 2018, after which the paper stagnated, and it is being put on arXiv now simply to broaden access.}
}

\date{Nov 14, 2018}
\author{ Aaditya Ramdas$^1$, Jianbo Chen$^2$,\\
 Martin J. Wainwright$^{2,3}$, Michael I. Jordan$^{2,3}$\\ 
 \texttt{aramdas@cmu.edu, jianbochen@berkeley.edu}\\
 \texttt{\{wainwrig,jordan\}@berkeley.edu}\\ 
 $^1$Department of Statistics and Data Science, Carnegie Mellon University\\
 Departments of Statistics$^2$ and EECS$^3$, UC Berkeley
 }

   
 

\begin{document}

\maketitle \thispagestyle{empty} \pagestyle{empty}

%%%%%%%%%%%%%%%%%%%%%%%%%%%%%%%%%%%%%%%%%%%%%%%%%%%%%%%%%%%%%%%%%%%%%%%%%%%%%%%%
\begin{abstract}
  This paper designs methods for decentralized multiple hypothesis testing on graphs that are equipped with provable guarantees on the false discovery rate (FDR). We consider the setting where distinct agents reside on the nodes of an undirected graph, and each agent possesses p-values corresponding to one or more hypotheses local to its node. Each agent must individually decide whether to reject one or more of its local hypotheses by only communicating with its neighbors, with the joint aim that the global FDR over the entire graph must be controlled at a predefined level. We propose a simple decentralized family of Query-Test-Exchange (QuTE) algorithms and prove that they can control FDR under independence or positive dependence of the p-values. Our algorithm reduces to the Benjamini-Hochberg (BH) algorithm when after graph-diameter rounds of communication, and to the Bonferroni procedure when no communication has occurred or the graph is empty. To avoid communicating real-valued p-values, we develop a quantized BH procedure, and extend it to a quantized QuTE procedure. QuTE works seamlessly in streaming data settings, where anytime-valid p-values may be continually updated at each node. Last, QuTE is robust to arbitrary dropping of packets, or a graph that changes at every step, making it particularly suitable to mobile sensor networks involving drones or other multi-agent systems. We study the power of our procedure using a simulation suite of different levels of connectivity and communication on a variety of graph structures, and also provide an illustrative real-world example.
\end{abstract}

\section{Introduction}
\label{sec:intro}

Research on decentralized detection and hypothesis testing has a long
history, dating back to seminal work from the 1980s
(e.g.,~\cite{tsitsiklis1984problems,Tenney81}), and with more recent
work motivated in particular by wireless sensor networks
(e.g.,~\cite{chamberland2003decentralized,alanyali2004distributed,olfati2006belief,Viswanathan97,Blum97}). A
variety of issues have been addressed, including
robustness~\cite{gul2017robust}, quantization
error~\cite{longo1990quantization}, the sequential nature of data
collection~\cite{fellouris2011decentralized}, battery/energy
management of the sensors~\cite{tarighati2016decentralized}, tolerance
of misbehaving nodes~\cite{soltanmohammadi2013decentralized},
nonparametric methods~\cite{Han90,NguWaiJor05}, and bandwidth
constraints~\cite{Luo05}.  However, these have focused on the ability to make a decision
about a single global hypothesis in a distributed fashion. In
contrast, this paper studies the testing of 
multiple hypotheses in a decentralized manner.

To be more specific, consider a graph with one agent at each node, and
suppose that each agent wishes to test one or more binary hypotheses
that are local to its node, with each hypothesis corresponding to the
presence or absence of some signal. Of this collection of hypotheses,
an unknown subset corresponds to true null hypotheses (absence of
signal), whereas the rest are false (presence of signal), and
should be rejected.  Apart from having access to the data and
hypotheses at its own node, each agent may also directly communicate
with neighboring nodes on the graph. Our setup is directly related to, yet complementary to, the works of~\cite{ermis2005adaptive,ray2007novel,xing2010weighted,ray2011false}.

For explicit context, one example of such an agent could be a
low-power sensor with the ability to perform local measurements and
simple computations. Then, the graph may represent a multitude of such
sensors deployed over a widespread area, such as a forest, building or
aircraft. Due to power constraints, the agent may communicate locally
with agents at neighboring nodes, but only infrequently communicate
with a distant centralized server. A rejected hypothesis corresponds
to a discovery of some sort, such as a detected anomaly or a deviation
from expected or recent measurements, and it is only in such cases
that a local node is allowed to communicate with the central server.

Each agent aims to test their local null hypotheses, and reject a
subset of nulls, proclaiming them to be false based on its local
neighborhood information. We assume that each agent can calculate a
$p$-value for each of its local hypotheses, which provides evidence
against the respective null. This paper is agnostic to the exact
methods that the node uses to calculate its local $p$-values; this
topic has received significant attention in past work, and can be done
in various ways.  Each agent must use the local $p$-values, and
possibly those of its neighbors, to decide which hypotheses to reject
locally. The challenge is that these \emph{local} decisions must be
made with the aim of controlling a \emph{global} error metric. In this
paper, the error metric is chosen as the overall false discovery rate
(FDR) across the whole graph, and we require that the FDR must be
controlled at a pre-defined level.

The false discovery rate (FDR) is an error metric that generalizes the
notion of type-1 error, or false positive error, to the setting of
multiple hypotheses. It is defined as the expected ratio of the number
of false discoveries to the total number of discoveries. It was first
proposed by Eklund~\cite{eklund1961massignifikansproblemet}, and
various heuristic methods were discussed by Eklund and
Seeger~\cite{eklund1965massignifikansanalys,seeger1968note}.  Its
contemporary usage has been popularized by the seminal paper of
Benjamini and Hochberg~\cite{BH95} who rigorously analyzed a procedure
that is now known as the Benjamini-Hochberg (BH) procedure.

Algorithms to control the FDR make sense in settings where type-1
errors are possibly more costly than type-2 errors.  Since such
an asymmetry arises naturally in many situations, the FDR has
nevertheless become one of the de-facto standard error metrics
employed across a wide range of applied scientific fields. Its most
common application is in a batched, centralized setting, where all the
$p$-values are pooled, and a decision is made by thresholding them at
some data-dependent level, such as the level determined by the
Benjamini-Hochberg (BH) procedure~\cite{BH95}.

In this paper, we develop a novel family of algorithms for
decentralized FDR control on undirected graphs. Both the setting and
the family of algorithms are new, to the best of our
knowledge. Building on recent theoretical advances, we prove that
global FDR over the whole graph is controlled under two types of
dependence assumptions between the $p$-values at the different nodes;
the setting of independent $p$-values is easiest, but a general type of
``positive dependence'' termed PRDS can also be handled.

A given instantiation of the QuTE algorithm is simple and scalable,
and works as follows. First, a node queries its neighbors for their
$p$-values. Second, it conducts a local test on the local neighborhood
$p$-values at an appropriate level determined by the ratio of the size
of its neighborhood to the size of the graph. Third, if a node thinks
that neighboring hypotheses should be rejected, it informs the
relevant neighbors. Lastly, each node rejects one or more of its local
hypotheses as long as either its local test suggested so, or if any of
its neighbors suggested so. We run simulations on a variety of graphs,
including Erd\"os-R\'enyi random graphs and planar grid graphs, to
provide some intuition about their achieved FDR and power. We also
give an illustrative example of possible real-world applications of
our setup and algorithm using a public Intel Lab dataset involving
low-power sensors taking various kinds of measurements in a large open
space.

Beyond guarantees on FDR control, our family of QuTE algorithms has
another appealing property. When the graph is either complete
(contains all edges) or a star, there exists at least one node that can
access all the information in the network, in which case our
algorithms reduce to the batch Benjamini-Hochberg procedure. When the
graph is empty (without edges), and each node tests only one
hypothesis without information about other nodes in the network, then
our algorithms essentially reduce to the conservative Bonferroni
procedure. Hence, one can view our algorithm as bridging the two
extremes, depending on how much information is available at each node.

The rest of this paper is organized as follows. \secref{back} presents
the problem setup, including definitions and assumptions. We formally
describe our algorithms and their associated theoretical guarantees in
\secref{algo}. We present the proof of our main theorem in
\appref{prooffdr} and \appref{proofpower}. We run a variety of
simulations in \secref{exp} and illustrate an application to real data
in \secref{real}. 
For efficient communication, a quantized version of the QuTE algorithm can be found in \secref{quantized-qute}.
Some extensions to anytime-valid p-values and evolving graphs can
be found in \secref{extensions}.

%%%%%%%%%%%%%%%%%%%%%%%%%%%%%%%%%%%%%%%%%%%%%%%%%%%%%%%%%%%%%%%%%%%%%%%%%%%%%%%%%%%%

\section{Background and problem setup}
\label{sec:back}

Consider a graph $G = (\V,\E)$, where each node $a \in \V$ is
associated with an agent.  A given agent $a$ may communicate directly
with any of its neighbors $b$, meaning nodes for which the edge
$(a,b)$ belongs to the edge set.  Associated with agent $a$ is a
collection $\HH_a = \{H_{a,i}\}_{i=1}^{n_a}$ of $n_a$ hypotheses to be
tested.  The sum $N = \sum_{a\in \V} n_a$ corresponds to the the total
number of hypotheses. Out of these, let $\nulls_a$ represent the
(unknown) subset of true null hypotheses at node $a$, and let $\nulls
= \cup_{a \in \V} \nulls_a$ represent the overall set of true null
hypotheses.  We assume that the agent at each node observes one
$p$-value corresponding to each hypothesis at that node, labeled
$P_{a,1},\dots,P_{a,n_a}$. It is understood that when a null
hypothesis is true, the corresponding $p$-value is stochastically
larger than the uniform distribution (super-uniform, for short),
meaning that
\begin{align}
\label{eq:superuniform}
 \text{ for all $H_{a,i} \in \nulls$, we have } \PP{P_{a,i} \leq u} \leq u \text{ ~ for all $u \in [0,1]$}.
\end{align}

Given observations of the $p$-values at its node, each agent may
communicate with its neighbors, and must then decide which subset of
its hypotheses to reject; each rejection is called a discovery, and
rejecting a true null hypothesis corresponds to a false discovery. Let
$R$ be the total number of discoveries, and $V$ be the total number of
false discoveries. Even though each agent makes a local decision with
only local information, we might still desire to control a global
measure of error. However, even in the centralized setting, there is
no single notion of \mbox{type-1} error for multiple testing problems;
we discuss two common variants below.

%%%%%%%%%%%%%%%%%%%%%%%%%%%%%%%%%%%%%%%%%%%%%%%%%%%%%%%%%%%%%%%%%%%%%%%%%%%%%%%%

\subsection{False Discovery Rate (FDR)}

One classical quantity is the family-wise error rate (FWER), which is
defined as $\PP{V > 0}$. A procedure that guarantees that $\text{FWER}
\leq \alpha$ might be desirable in applications where a false
discovery is extremely costly, and must be avoided at all costs. One
simple way to control the FWER is the Bonferroni procedure, which
rejects every $p$-value that is smaller than $\alpha/N$; the FWER
guarantee follows by an application of the union bound.

However, in many practical settings involving even a moderately large
number $N$ of hypotheses, the FWER metric is too stringent and the
associated procedures (like Bonferroni and its variants) have very low
power, making very few discoveries for a given level of FWER control.
This issue led Eklund~\cite{eklund1961massignifikansproblemet}, in
collaboration with Seeger~\cite{eklund1965massignifikansanalys}, and
more recently Benjamini and Hochberg~\cite{BH95} to propose an
alternative notion of error, called the false discovery rate (FDR),
which we now define and adopt in this work.  We adopt the convenient
notation $\dotfrac{a}{b} = \frac{a}{\max{b,1}}$, and define the false
discovery rate (FDR) to be
\begin{align*}
\small \fdr = \EE{\dotfrac{V}{R}}.
\end{align*}
As suggested by the name, it corresponds to the expected proportion of
false discoveries relative to the total number of discoveries.  On
average, a procedure with $\fdr$ controlled at level $\alpha$ will
return a collection of discoveries of which at most an
$\alpha$-fraction are false.

%%%%%%%%%%%%%%%%%%%%%%%%%%%%%%%%%%%%%%%%%%%%%%%%%%%%%%%%%%%%%%%%%%%%%%%%%%%%%

\subsection{Centralized algorithms for FDR control}

All existing algorithms for FDR control have been designed for the
centralized setting, where a single agent has access to all $N$
$p$-values. In such a setting, if $P_{(j)}$ is the $j$-th smallest
$p$-value, then the Benjamini-Hochberg (BH) algorithm~\cite{BH95}
rejects the hypotheses corresponding to the smallest $\kh$ $p$-values
$P_{(1)},\dots,P_{(\kh)}$, where $\kh$ is a data-dependent integer
chosen by the rule
\begin{align*}
\kh^{BH} = \max \big\{ k \mid P_{(k)} \leq \alpha \frac{k}{N} \big\}.
\end{align*}
This procedure guarantees that the FDR is controlled at level $\alpha$
when the $p$-values are independent~\cite{BH95}. Equivalently, the
$\bh$ procedure rejects all $p$-values that are at most
$\alpha \frac{\kh^{BH}}{N}$.

Benjamini and Yekutieli~\cite{BY01} proved that the BH algorithm also
guarantees FDR control when the $p$-values satisfy a general notion of
positive dependence called positive regression dependence on a subset
(PRDS). Since our results also hold under this condition, we define it
formally now.

For a pair of vectors $x, y \in [0,1]^N$, the notation $x \preceq y$
means that $x \leq y$ in the orthant ordering, i.e., $x_j \leq y_j$
for all $j \in \{1, \dots, N\}$.
% \begin{definition}[Nonincreasing sets and functions]
We say that a function $f: [0,1]^N \mapsto \mathbb{R}^+$ is
\emph{nonincreasing}, if $x \preceq y$ implies $f(x) \geq f(y)$.
Similarly, we say that a subset $\Dset$ of $[0,1]^n$ is
\emph{nondecreasing} if $x \in \Dset$ implies $y \in \Dset$ for all $y
\succeq x$.
% \end{definition}

\begin{definition}[PRDS, positive dependence]
\label{ass:PRDS}
For any nondecreasing set $ \Dset \subseteq [0,1]^N$ and $H_{a,j} \in
\nulls$, the function $t~\mapsto~\PPst{P\in \Dset}{P_{a,j} \leq t}$ is
nondecreasing on $(0,1]$.
\end{definition}
% When the PRDS condition is met for a particular null $P_i$, we then
% say that \emph{$P$ is PRDS on $P_i$}.  The original positive
% regression dependence assumption~\cite{lehmann1966some} as well as
% the PRDS assumption~\cite{BY01} both had $P_i = t$ instead of $P_i
% \leq t$ in the definition, but one can prove that both conditions
% are essentially equivalent.

The PRDS condition holds trivially if the $p$-values are independent,
but also allows for some amount of ``positive'' dependence. Let us
consider a simple example to provide some intuition.  Let $Z =
(Z_1,\ldots,Z_N)$ be a multivariate Gaussian vector with covariance
matrix $\Sigma$; the null components correspond to Gaussian variables
with zero mean.  Letting $\Phi$ be the CDF of a standard Gaussian, the
vector $P=(\Phi(Z_1),\dots,\Phi(Z_N))$ is PRDS (on every null index)
if and only if all entries of the covariance matrix $\Sigma$ are
non-negative.  See the paper~\cite{BY01} for other examples of this
type.
% It should be noted the PRDS assumption is closely
% related to the assumption of multivariate total positivity of order 2
% (MTP2)~\cite{karlin1980classes}. The latter condition is sometimes
% also called log-supermodularity, and arises in lattice theory,
% percolation theory and theoretical computer science, because it
% implies the famous Fortuin-Kasteleyn-Ginibre (FKG) inequality. Since
% MTP2 is known to imply PRDS, all the results in this paper that hold
% under PRDS also immediately also hold under MTP2.

The BH algorithm has been extended to a variety of settings;
see~\cite{pf+} for a survey.  In this paper, we consider the new
setting of decentralized multiple testing, for which there are no
known algorithms or guarantees for FDR control.

%%%%%%%%%%%%%%%%%%%%%%%%%%%%%%%%%%%%%%%%%%%%%%%%%%%%%%%%%%%%%%%%%%%%%%%%%%%%%%%%

\subsection{Communication Protocol}

In engineering applications such as low-power wireless sensor
networks, often there is no central agent that has access to all
$p$-values. Indeed, while it might be possible for each sensor to
communicate with a central server, such communication could consume
signficant power if the distances involved are large. It would make
more sense for each sensor to communicate with a central server only
if something anamolous is happening at or around that node,
corresponding to a (rare) rejected hypothesis.

A reasonable model is that each agent can communicate only with its
neighbors. In this work, we assume that noiseless real numbers (like
$p$-values) can be transmitted without error or quantization.
% This allows us to concentrate on designing algorithms where the emphasis is
% on achieving decentralized FDR control. Future work may possibly
% consider the effects of errors or quantization on the FDR guarantees.
Except for possible communication between the agents, the problem we
consider is static. This means that the graph structure, the
hypotheses at each node, and their associated $p$-values, do not
change with time. In other words, we consider a snapshot of network at
one point in time, leaving extensions for future work.

Decentralized testing is, almost by definition, harder than
centralized testing. Indeed, it should be intuitively clear that among
all algorithms that guarantee FDR control, a centralized one will make
more discoveries, because it has access to more information. Hence,
the price that we will pay by restricting communication, while still
demanding FDR control, will be in the achieved power of the
algorithms, which is their ability to make true discoveries.

%%%%%%%%%%%%%%%%%%%%%%%%%%%%%%%%%%%%%%%%%%%%%%%%%%%%%%%%%%%%%%%%%%%%%%%%%%%%

\section{The QuTE Algorithm}
\label{sec:algo}

For simplicity, we first describe a single-step Query-Test-Exchange
(QuTE) algorithm. If more communication is feasible, then the
extension to a multi-step QuTE algorithm is straightforward, and
presented in a later subsection.

We assume that each node has a unique ID, and that the agent knows
this ID. Each agent $a$ fixes an order of the hypotheses (and
corresponding $p$-values) at its node, and forms a vector $\vec{P}_a$
with an ordered list of $p$-values at node $a$.  The single-step QuTE
algorithm proceeds as follows:
\begin{itemize}
\item \textbf{Query}: Each agent queries its neighbors for the vectors
  that they are in possession of. Each agent receives an ordered
  vector of $p$-values from every neighbor, and keeps track of which
  neighbor each $p$-value came from.
\item \textbf{Test}: Let $S_a$ be the set $p$-values now in possession
  of agent $a$. Agent $a$ runs the BH procedure on its $|S_a|$
  $p$-values, at an adjusted target FDR of $\alpha_{a} := \alpha
  \frac{|S_a|}{N}$. Equivalently, agent $a$ assumes that all $p$-values it
  does not know equal 1, and runs the BH procedure on all $N$ $p$-values at level $\alpha$.
  Denoting the number of rejections at node $a$ by
  $\kh_a$, it is natural that these $\kh_a$ rejections may include
  some of agent $a$'s own hypotheses, and some from its neighbors.
\item \textbf{Exchange}: All agents exchange their rejection decisions
  with their neighbors by sending back an ordered indicator vector,
  informing the neighbors to reject every hypothesis that has its
  indicator location set to unity.
\end{itemize}
At the end of a single pass of these three steps, each agent then
rejects any hypothesis that it has been informed that it should
reject. Hence, an agent rejects any local hypothesis rejected by its
own test, or by the tests of any of its neighbors. This algorithm
comes with the following guarantee.

\vspace{0.1in}
\begin{theorem}\label{thm:fdr}
  Suppose that $p$-values are independent or positively dependent.
  Then for any graph topology, the aforementioned QuTE algorithm achieves FDR control
  at level $\alpha \frac{|\nulls|}{N}$.
\end{theorem}
\vspace{0.1in}

See~\appref{prooffdr} for the proof of~\thmref{fdr}.  The algorithm is
seemingly quite liberal with its rejections, in the sense that when
determining whether a hypothesis should be rejected or not, each agent
does not take a majority vote amongst its neighbors, but rejects
whenever even a single neighbor thinks it should be
rejected. Nevertheless, it still manages to guarantee FDR control on
any graph.

%%%%%%%%%%%%%%%%%%%%%%%%%%%%%%%%%%%%%%%%%%%%%%%%%%%%%%%%%%%%%%%%%%%%%%%%%%%%%%%%%%%%%

\subsection{Dependence on graph topology}

Informally speaking, more edges in the graph results in better
connectivity, more information per agent, and higher power. 
We can precisely characterize what the QuTE algorithm does on
some special graph structures below:
\begin{itemize}
\item When the graph has no edges, and each agent has only one
  hypothesis, then QuTE reduces to the Bonferroni
  correction, where a hypothesis is rejected if and only if its
  $p$-value is at most $\alpha/N$.
\item When there exists at least one node which is connected to all
  others, like in a star graph or a clique, then after the query step,
  this agent possesses all the $p$-values, and then  QuTE
  reduces to the classical BH algorithm.
\end{itemize}

The price that one pays for limited information is power. We can make
this somewhat precise, by introducing a partial ordering over
graphs. For any two graphs on the same node set, we say that $G_1
\sqsupset G_2$ whenever $E_1 \supset E_2$. The relation $\sqsupset$
defines a partial ordering over such graphs. Fix the number of nodes,
and also the hypotheses and $p$-values at every node, and let
$\mathcal{R}(G)$ denote the set of rejections made by the QuTE
algorithm on graph $G$, and let $R(G) = |\mathcal{R}(G)|$. We then
have the following result.

\vspace{0.1in}
\begin{proposition}
  \label{prop:power}
The number of rejections made by the QuTE algorithm is nondecreasing
with respect to $\sqsupset$, meaning that conditioning on the observed
$p$-values, we have
\begin{align*}
G_1 \sqsupset G_2 \implies \mathcal{R}(G_1) \supseteq
\mathcal{R}(G_2).
\end{align*}
\end{proposition}
\vspace{0.1in}
\noindent {\bf{Remark:}}
By taking absolute values, and then expectations over the randomness
in $p$-values, we also conclude that $G_1 \sqsupset G_2$ implies that
$\EE{R(G_1)} \geq \EE{R(G_2)}$, meaning that the power of QuTE is also
nondecreasing with respect to $\sqsupset$.

See \appref{proofpower} for the proof of \propref{power}. An immediate
consequence of this proposition is that the power of QuTE can never
exceed the power of BH. This is simply because the clique dominates
all graphs with respect to $\sqsupset$, and QuTE reduces to BH on the
clique.  Some other effects of graph topology are explored in the
simulations.

%%%%%%%%%%%%%%%%%%%%%%%%%%%%%%%%%%%%%%%%%%%%%%%%%%%%%%%%%%%%%%%%%%%%%%%%%%%%%%%%%
\subsection{Extension to multi-step QuTE algorithms} 

The essential ingredient in the proof of \propref{power} is the
following fact: the more $p$-values to which a node has access, the more
rejections its local test can make. This fact opens up the possibility
of having $c>1$ rounds of querying before conducting the local test,
followed by $c >1$ rounds of exchanging the results after testing. We
call this the multi-step QuTE algorithm.

To be more explicit, multiple rounds of communication allow an agent
to access the $p$-values at nodes that are not immediate
neighbors. Indeed, with $c$ rounds of communication, it is
straightforward that a node can access $p$-values of all nodes that are
at a distance $\leq c$ away from it on the graph. After performing its
local test, $c$ rounds of exchanging information can once again
propagate the results of its local test to nodes that are at a distance
$\leq c$ away.

\vspace{0.1in}
\begin{theorem}\label{thm:fdr2}
If the $p$-values are independent or positively dependent, then the
multi-step QuTE algorithm with $c \geq 0$ rounds of communication
guarantees that $\fdr \leq \alpha \frac{|\nulls|}{N}$.
\end{theorem}
\vspace{0.1in}

The proof of~\thmref{fdr2} is very similar to the proof
of~\thmref{fdr}, so we do not reproduce it. 
We can precisely characterize the QuTE algorithm at the two extremes of communication:
\begin{itemize}
\item When $c=0$ rounds of communication are allowed, 
and each agent has only one hypothesis, then QuTE reduces to the Bonferroni
  correction, where a hypothesis is rejected if and only if its
  $p$-value is at most $\alpha/N$.
\item If $d$ is the diameter of the graph, then $\lceil d/2 \rceil$
rounds of communication suffice for QuTE to achieve same power as a
centralized BH procedure (and further communication makes no difference). 
\end{itemize}

Note that the above stated bound is a worst-case upper bound on how much communication suffices, and one can often achieve the same power as a centralized algorithm with fewer than $\lceil d/2 \rceil$ rounds. 

% Indeed, it is not necessary for all nodes to have information about all other nodes: 
% it suffices for a single central node, like in a star graph, to access all $p$-values.

%%%%%%%%%%%%%%%%%%%%%%%%%%%%%%%%%%%%%%%%%%%%%%%%%%%%%%%%%%%%%%%%%%%%%%%%%%%%%%%%%%

\section{Simulations}
\label{sec:exp}

\begin{figure*}[h!]
  \centering \subcaptionbox{}[.3\linewidth][c]{%
    \includegraphics[width=\linewidth]{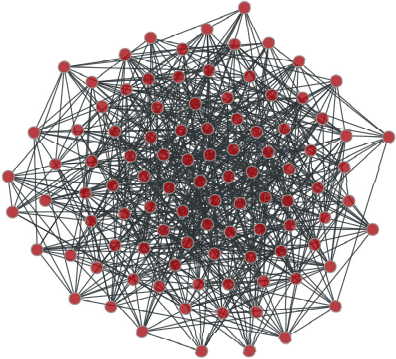}}\quad
  \subcaptionbox{}[.3\linewidth][c]{%
    \includegraphics[width=\linewidth]{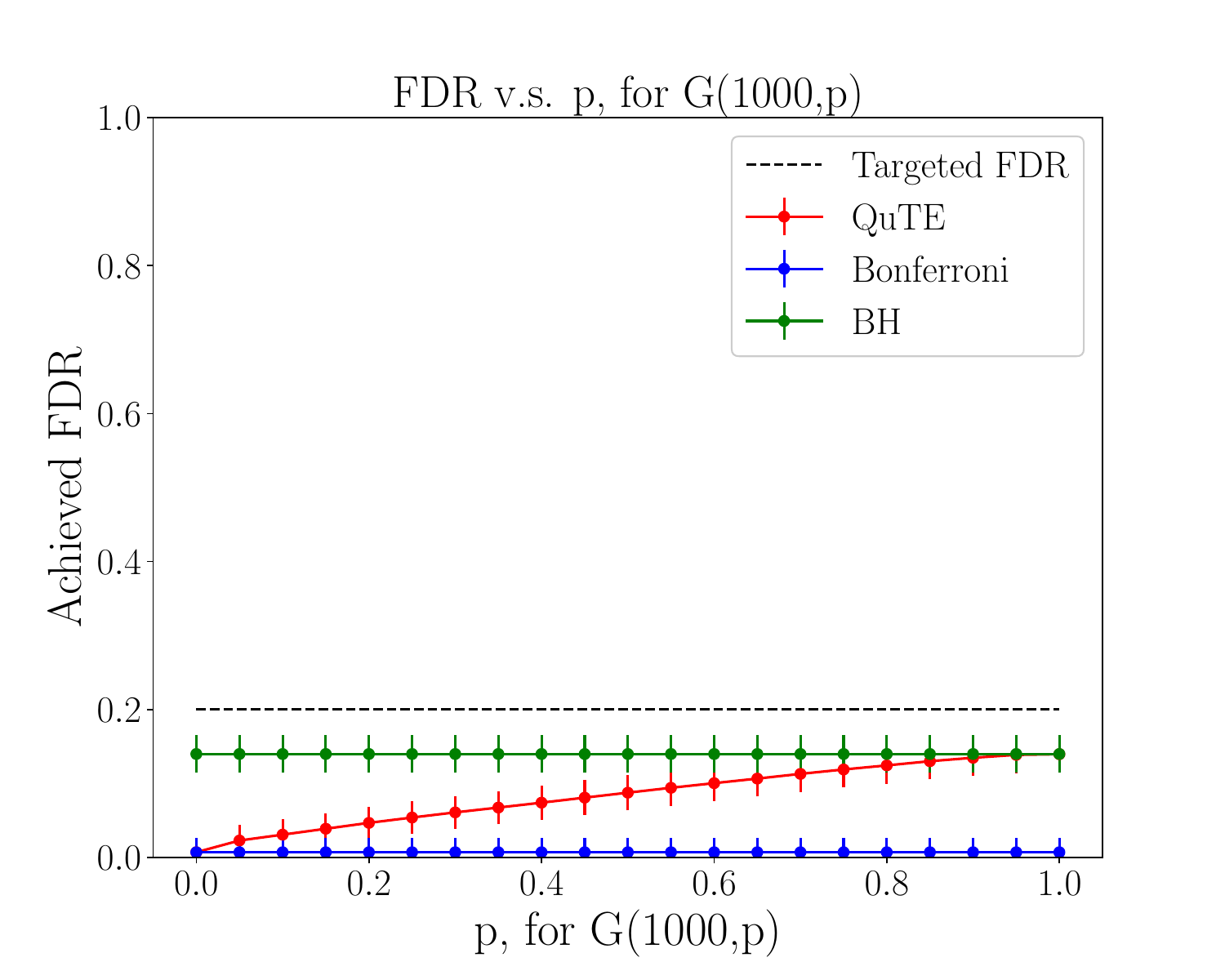}}\quad
  \subcaptionbox{}[.3\linewidth][c]{%
    \includegraphics[width=\linewidth]{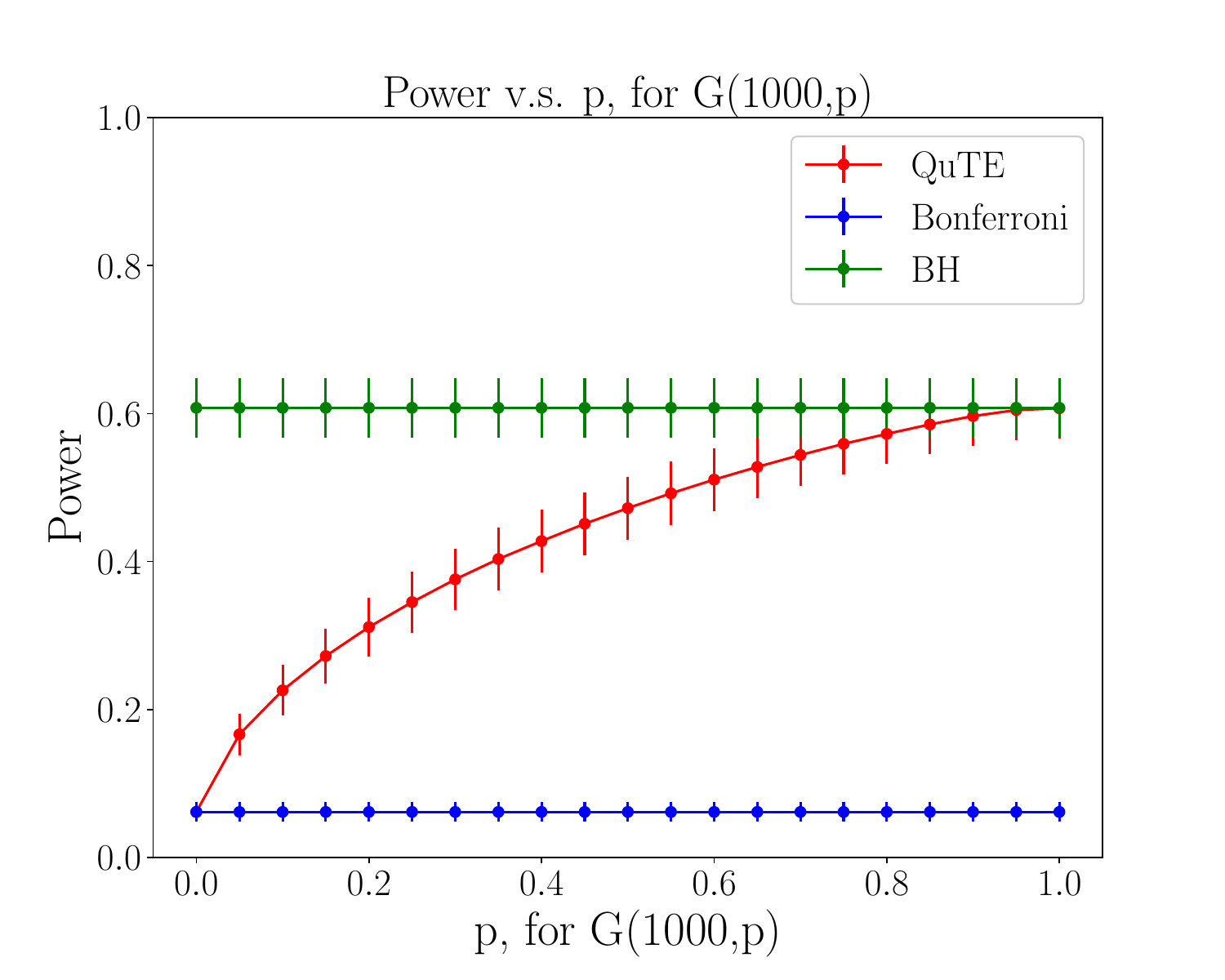}}
\caption{(a) Illustration of an
  Erd\"os-R\'enyi random graph $G(100,0.15)$. (b) Plot of
  FDR versus edge probability $p$.  (c) Plot of power versus edge
  probability $p$.  Panels (b) and (c) show results for the
  QuTE, BH and Bonferroni procedures applied to the random graph model
  $G(1000, p)$, with target FDR $\alpha=0.2$, signal
  level $\mu=2$ and the null-proportion $\pi_0=0.7$ .}
\label{fig:gnp}
\end{figure*}  

\begin{figure*}
  \subcaptionbox{}[.3\linewidth][c]{%
    \includegraphics[width=\linewidth]{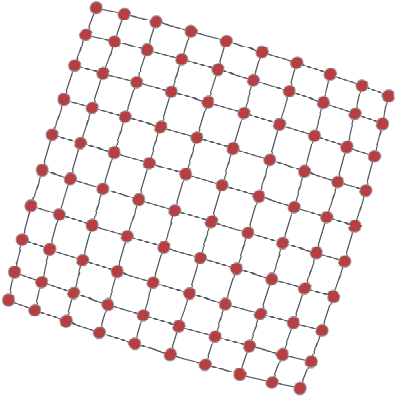}}\quad
  \subcaptionbox{}[.3\linewidth][c]{%
    \includegraphics[width=\linewidth]{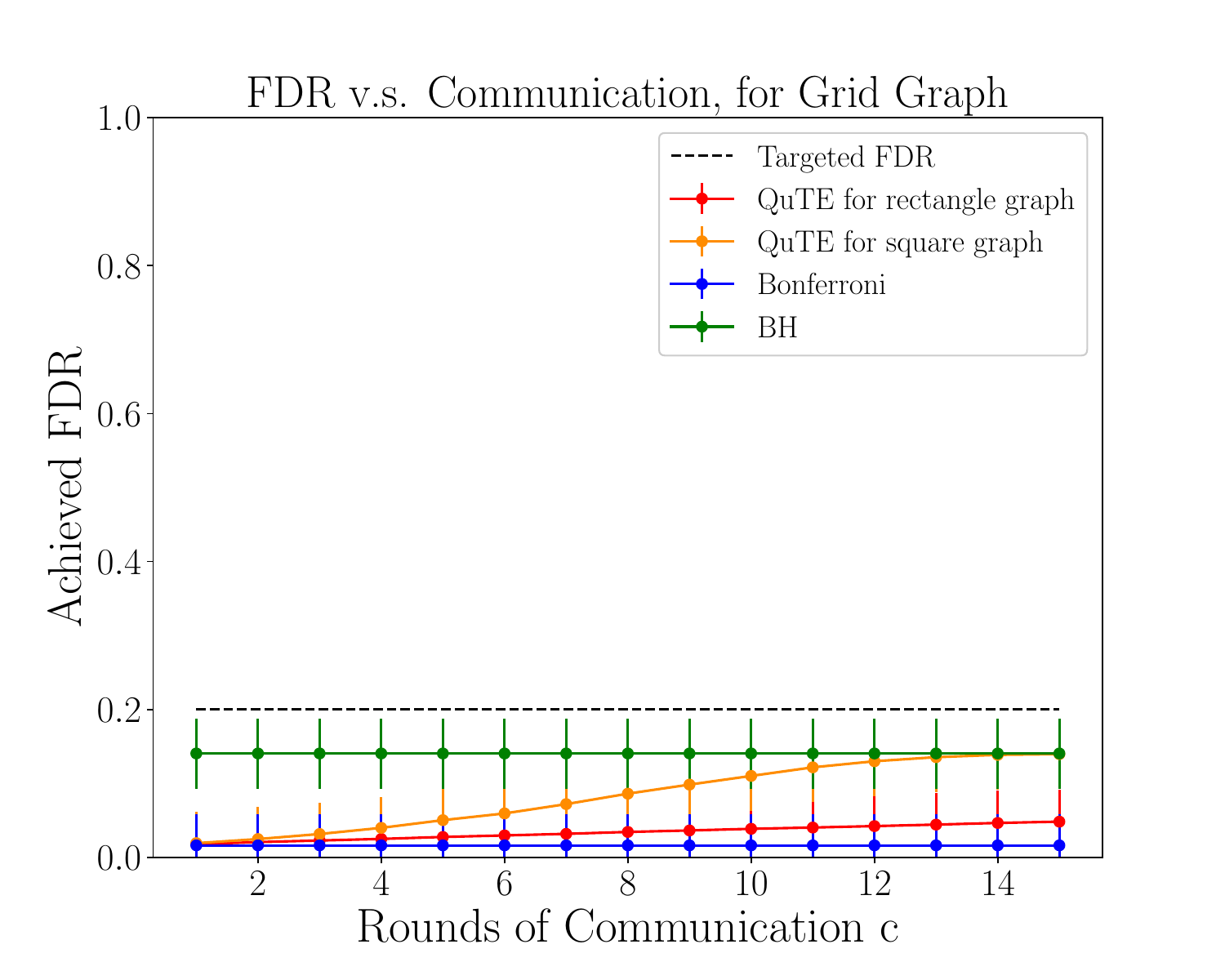}}\quad
  \subcaptionbox{}[.3\linewidth][c]{%
    \includegraphics[width=\linewidth]{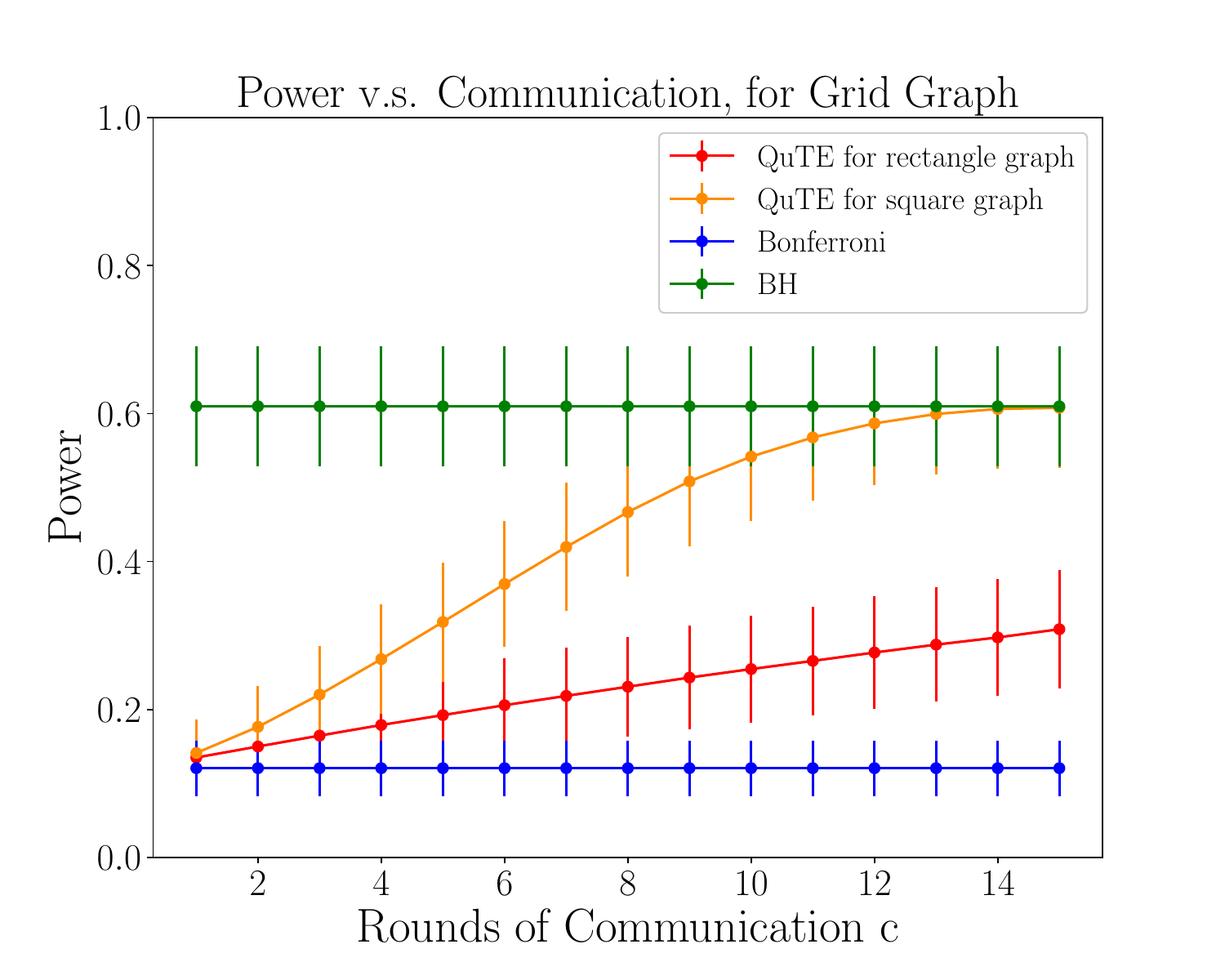}}
  \caption{(a) Illustration of a $10\times10$ grid graph.  Panels (b) and (c) display FDR and power versus rounds of communcation
     on a $16\times
    16$ grid graph and $2\times 128$ grid graph, for multi-step QuTE for a BH
    procedure, and Bonferroni procedure, with target FDR $\alpha=0.2$, signal level $\mu=2$ and the null proportion $\pi_0=0.7$.}
\label{fig:grid}
\end{figure*}

We now examine the behavior of QuTE on two different
types of graphs, Erd\"os-R\'enyi random graphs and grid graphs. In both settings, each node tests a single hypothesis, with a $p$-value independently generated
from the following model:
\begin{align}
  \label{$p$-value-model}
  \small
X\sim \mu + \mathcal{N}(0,1); \text{ $p$-value } P = 1-\Phi(X),
\end{align}
where $\Phi$ is the standard Gaussian CDF, with $\mu=0$ for nulls and
$\mu>0$ for alternatives. In both experiments, we compare the performance of QuTE with
the BH method and the Bonferroni correction, fixing the target FDR as $\alpha=0.2$, signal
strength $\mu=2$ and null-proportion $\pi_0=\frac{|\nulls|}{N}=0.7$. Code for reproducing empirical results is available
online. \footnote{\url{https://github.com/Jianbo-Lab/QuTE/}}

%%%%%%%%%%%%%%%%%%%%%%%%%%%%%%%%%%%%%%%%%%%%%%%%%%%%%%%%%%%%%%%%%%%%%%%%%%%%%%%%%%%%%

\subsection{Erd\"os-R\'enyi Random Graphs}

We first study the effects of increasing edge density for
 Erd\"os-R\'enyi random graphs $G(N,p)$. 
 In this model, an edge between any two distinct nodes is included in the graph
with probability $p$ independently, as illustrated in
\figref{gnp}(a). In our experiments, we set $N=1000$, varying the edge
probability $p$ from $0$ to $1$ in steps of $0.05$. We then randomly generate $1000$ $p$-values
from model \eqref{$p$-value-model}, each randomly assigned to one node
of the graph.

Results are shown
in panels (b) and (c) of \figref{gnp}.  We repeat our
experiments $100$ times, and the error bars are taken to be the
empirical standard deviations. One can observe that both the achieved
FDR and power increase with the edge probability $p$. In particular,
recall that QuTE reduces to Bonferroni when there is no edge ($p=0$)
and reduces to BH test when the graph is complete ($p=1$).

%%%%%%%%%%%%%%%%%%%%%%%%%%%%%%%%%%%%%%%%%%%%%%%%%%%%%%%%%%%%%%%%%%%%%%%%%%%%%%%%

\subsection{Grid Graphs}

We now study the effect of
communication and graph structure on power, within the family of
two-dimensional grids, that is, a planar graph whose vertices
correspond to points with integer coordinates $(x,y)$, with $x=1,\dots,m$ and
$y=1,\dots,n$. Edges exist between any two points with unit
distance,  as exemplified in
\figref{grid}(a). Variants of such graphs may be encountered in
environmental applications, where sensors are placed over a forest or lake. 
We consider a square grid with $m=n=16$ and a rectangular grid with
$m=2,n=128$.  In either case, $N = m\times n = 256$ $p$-values are first
randomly generated from model \eqref{$p$-value-model} and then randomly
assigned to the nodes. We vary the number of the rounds
of communication $c$ in multi-step QuTE on the two grids.

Results as the parameter $c$ varies are shown in \figref{grid} (b, c).
For each choice of $c$, we repeat the simulation $1000$ times, and the
error bars are again taken to be the empirical standard deviations. In
both grids, the achieved FDR and the power increase with the rounds of
communication $c$. The square grid graph always achieves a higher FDR
and power when $c$ is large, as each node can effectively get
information from a larger number of nodes. Multi-step QuTE achieves
comparable power as BH on the square grid graph when the number of the
rounds of communication is close to $c=16$, when the center node can
reach all of the nodes on the graph.

%%%%%%%%%%%%%%%%%%%%%%%%%%%%%%%%%%%%%%%%%%%%%%%%%%%%%%%%%%%%%%%%%%%%%%%%%%%%%%%%

\section{An Illustrative Example on Real Data} \label{sec:real}

\begin{figure}[h!]
\centering
    \includegraphics[width=0.6\linewidth]{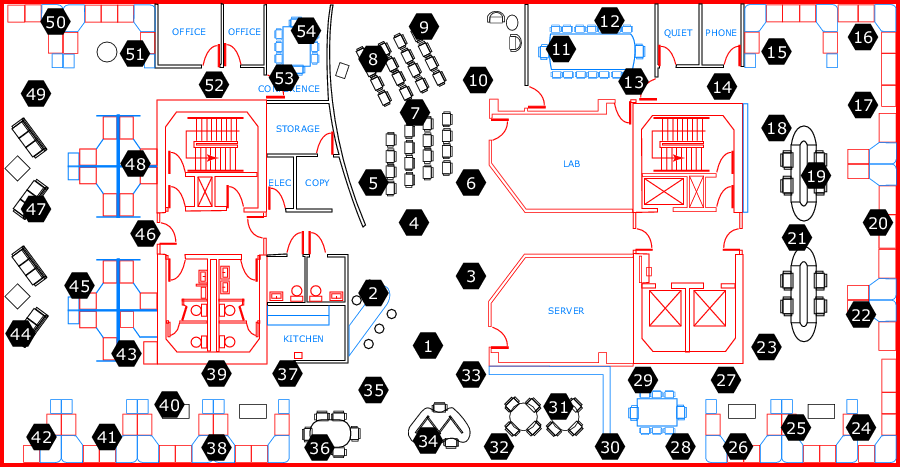}
    \caption{Arrangement of sensors in the lab~\cite{labdata}.}
    \label{fig:lab}
\end{figure} 

\begin{figure*}[t]
\centering
  \subcaptionbox{Threshold $\gamma=0.1$}[.29\linewidth][c]{%
    \includegraphics[width=\linewidth]{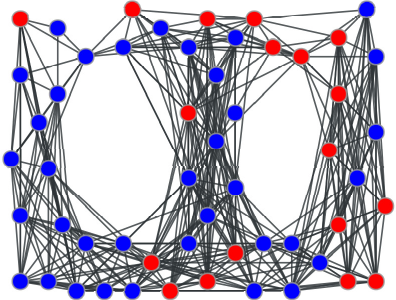}}\quad
  \subcaptionbox{Threshold $\gamma=0.3$}[.29\linewidth][c]{%
    \includegraphics[width=\linewidth]{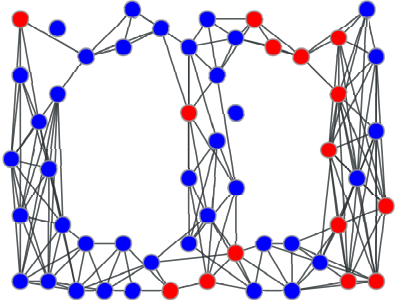}}\quad
  \subcaptionbox{Threshold $\gamma=0.5$}[.29\linewidth][c]{%
    \includegraphics[width=\linewidth]{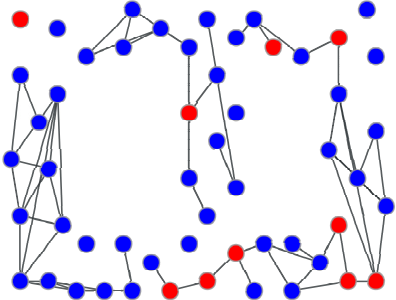}}
  \caption{The three figures show the result of test at a common
    sample ($100$ epochs). The graphs are specified by choosing the
    threshold on communication probability to be $0.1,0.3,0.5$ for
    (a),(b),(c) respectively. The nodes in red are rejected.}
  \label{fig:realdata}  
\end{figure*} 

We now provide an illustrative example for potential applications of
QuTE to sensor networks. We use the publicly available Intel Lab
dataset~\cite{labdata}, which contains data collected from $54$
sensors deployed in the Intel Berkeley Research Lab. The distribution
of sensors is shown in \figref{lab}. Mica2Dot sensors with weather
boards collected timestamped topology information, along with
humidity, temperature, light and voltage values once every $31$
seconds for $36$ days. Here, we only use the temperature measurement
for simplicity, but note that our theory can handle the case when
there are several hypotheses per node, possibly with positively
dependent $p$-values.  Each sensor can communicate with another sensor
with certain probability, which is estimated by averaging connectivity
data over time. This communication probability is provided in the
dataset \cite{labdata}. For simplicity, we threshold the probability
and put an edge between two sensors if their communication probability
with each other is larger than a threshold $\gamma \in \{0.1,0.3,
0.5\}$.

\paragraph{Data preprocessing}

The temperature data has a slowly increasing trend. We detrend
the data for each sensor by first fitting a regression line and then subtracting the
least squares fit. To reduce noise, we
average over every consecutive $100$ epochs for each sensor to get a
sample. That is, one sample has $54$ average temperature measurements
from respective sensors within a consecutive $100$ epochs.

\paragraph{Hypotheses and $p$-values}

At each node,
the null hypothesis states that the current temperature at the node is
normal over one randomly chosen time sample ($100$ epochs).

Define the empirical distribution of the temperature of the first
$500$ samples at sensor $a$ as $\hat F_{n,a}$.  Then we define the
$p$-value at sensor $a$ as
% \begin{align}
$P_a = 2 \min(1-\hat F_{n,a}(t_a),\hat F_{n,a}(t_a)),$
% \end{align}
where $t_a$ is the temperature of sensor $a$ in the current sample. We
adopt the non-asymptotic calculation of $p$-values instead of using
standard student t-tests because the empirical distribution has a
heavy tail and is not asymptotically normal. In any case, this choice
was made for illustration, and our method is agnostic to the exact
method used to construct $p$-values, as long as they satisfy condition \eqref{eq:superuniform}. As temperature
measurements at the same time are positively correlated, the $p$-values
are also positively correlated and we are justified in using QuTE to
control FDR.

\paragraph{Results}

The three subplots of \figref{realdata} display the results for three
graphs formed by thresholding communication probabilities at $\gamma
\in \{0.1,0.3,0.5\}$. The nodes in red are rejected, and naturally,
there are more rejections on average with increasing density of
edges. Given that QuTE provably controls the FDR under positive
dependence, one can expect a large number of the rejected nodes to be
true discoveries of an anomalous temperature at the current snapshot
of time.

\section{The quantized Benjamini-Hochberg and QuTE procedures}
\label{sec:quantized-qute}

The aforementioned QuTE algorithm transmits real-valued $p$-values along the edges of the graph at each step. In practice, it is natural to avoid transmitting real-valued observations. Hence, we next introduce the notion of a $p$-rank which are integers between $1$ and $N$, and develop a ``quantized BH'' procedure that is written in terms of $p$-ranks instead of $p$-values, and makes the same decisions as the original one. This allows one to %quantize the QuTE procedure via $p$-ranks, and hence only 
transmit integers using at most $\lceil \log (N) \rceil$ bits per $p$-rank.

We define the $p$-rank associated with a $p$-value $P_i$ as
\[
\prank_i = \min(\lceil P_i N/\alpha \rceil, N).
\]
Since the BH procedure cannot reject $p$-values larger than $\alpha$, the truncation at $N$ does not matter for power.
It is important to note that $p$-ranks are always integers between $1$ and $N$, and %they are not unique. Indeed, 
different $p$-values could be mapped to the same $p$-ranks, since
\[
\prank_i = s ~\iff~  (s-1)\alpha/N  < P_i \leq s\alpha/N,
\]
and there might be several $p$-values in the same range. 

We define the quantized BH algorithm (or BH$^Q$ for short), which works only with $p$-ranks, as follows. It first sorts the $p$-ranks, breaking ties arbitrarily. Calling the $i$-th sorted $p$-rank as $\prank_{(i)}$, we define
\[
\kh^Q := \max\{k: \prank_{(k)} \leq k\}.
\]
Then BH$^Q$ rejects the first $\kh^Q$ of the sorted hypotheses, meaning it rejects every hypothesis such that $\prank_{(k)} \leq \kh^Q$. 

The quantized QuTE procedure is defined analagously --- in every Query and Exchange round, only quantized $p$-values are transmitted, and in each Test round, each node runs BH$^Q$ at level $\alpha$ on its $p$-values (while assuming all unknown $p$-values equal 1). Then, we have the following result.

\begin{theorem}\label{thm:qbh}
The quantized BH procedure makes exactly the same discoveries as the BH procedure. As an immediate consequence, the quantize QuTE procedure makes exactly the same discoveries as the QuTE procedure.
\end{theorem}

A simple proof is provided in \appref{proofqbh}.

\section{Some extensions}
\label{sec:extensions}

\paragraph{Anytime-valid $p$-values for sequential QuTE.}
For simplicity of notation, below we assume that each node is testing just one hypothesis (this is by no means a restriction, and the results go through seamlessly with multiple hypotheses per node).

The presented QuTE algorithm assumed that the $p$-values $P_a$ were fixed, and the only dynamism involved was to do with the communication protocol, where the passage of time only affected how much information each node knew about the rest of the graph.
Here, we argue that QuTE can be easily extended to settings involving sequential testing, where data streams in over time, and hence the $p$-values $P_{a,t}$ change over time as more data is accumulated. 

 In the non-sequential setting, null $p$-values $P_a$ are constructed to be superuniform when the null hypothesis is true, as seen in condition \eqref{eq:superuniform}. The natural generalization of this condition to the sequential setting is that the $p$-values $P_{a,t}$ are superuniform at any (possibly random) time:

\begin{definition}[Anytime-valid $p$-value]
We call a sequence of $p$-values $(P_t)_{t \geq 1}$ testing a null hypothesis $H$ as an anytime-valid $p$-value if
\begin{align}
\label{eq:alwaysvalid}
\text{if $H$ is null, then we have } (\star)~\PP{\exists t \in \N :  P_{t} \leq u} \leq u \text{ ~ for all $u \in [0,1]$}.
\end{align}
Equivalently, we may replace the condition $(\star)$ with the condition $\PP{P_\tau \leq u} \leq u$ for any stopping time $\tau$.
\end{definition}

Without loss of generality, anytime-valid $p$-values are decreasing sequences: if $(P_t)_{t \geq 1}$ is anytime-valid, then so is $(\min_{s\leq t} P_s)$ --- the definition insists that under the null they will never drop below $\alpha$, with probability at least $1-\alpha$. The QuTE procedure can be easily seen to work seamlessly with anytime-valid $p$-values instead of fixed $p$-values. In this case nodes work with potentially stale versions of anytime-valid $p$-values (meaning that by the time a version of $P_{a,t}$ reaches node $b$ at time $t' > t$, $P_{a,t'}$ is already smaller unbeknownst to node $b$). Then, the QuTE procedure will simply produce a subset of discoveries of what a centralized BH procedure would have produced (and a subset of discoveries that a QuTE procedure with nearly instantaneous communication would), but the FDR will be controlled under the same conditions as QuTE itself.

\begin{figure*}[h!]
\centering
    \includegraphics[width=0.8\linewidth]{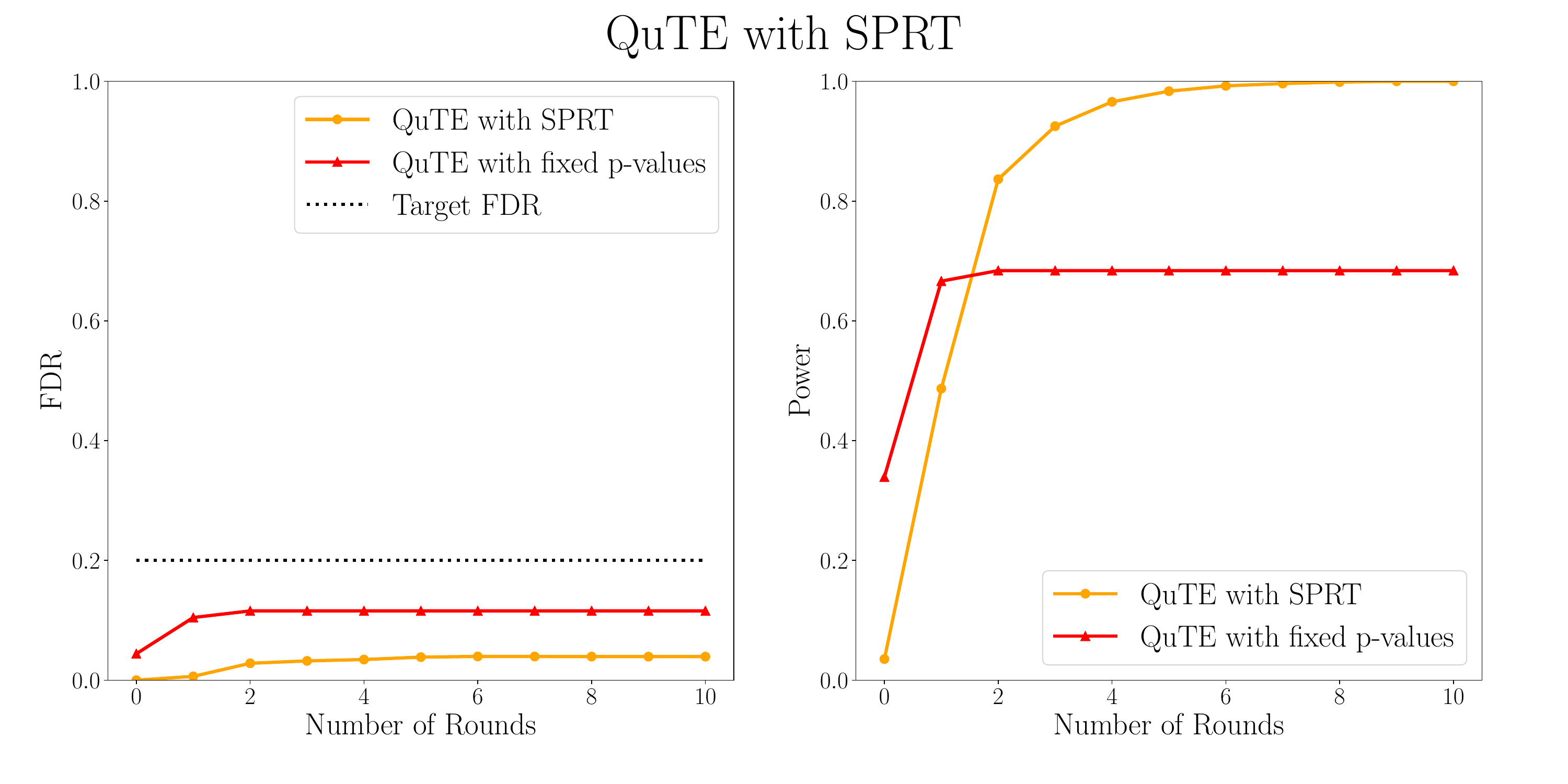}
      \caption{This figure demonstrates that QuTE with anytime-valid p-values (in this case, computed from a stream of incoming data using the SPRT) can be more powerful than using classical p-values (calculated from a fixed batch of data).}
      \label{fig:SPRT}  
\end{figure*}

% The equivalence of the aforementioned conditions was proved in ????.

\paragraph{Automatic robustness to dropped packets and changing graphs.}

We note that QuTE, even with anytime-valid $p$-values, is automatically robust to dropped packets and graphs that change topology with time (like a set of drones whose relative positions, and thus interconnectivities, keep varying).

As mentioned above, dropped packets can only cause QuTE to become more conservative, but not violate FDR control. The reason is simply because every node, when it does not know a particular $p$-value, just replaces it with $1$ (or whatever it knew last about that $p$-value) for its local multiple test. Every graph can be seen simply as the full graph, but with certain packets (along the missing edges) being systematically dropped at each step. A changing graph can thus be viewed as the full graph, but which packets get dropped changes at each step: this cannot change the FDR control guarantee, only potentially make it take longer for some node to learn the p-value of some other node, thus delaying a potential rejection and hurting power.

\begin{figure*}[h!]
\centering
    \includegraphics[width=0.75\linewidth]{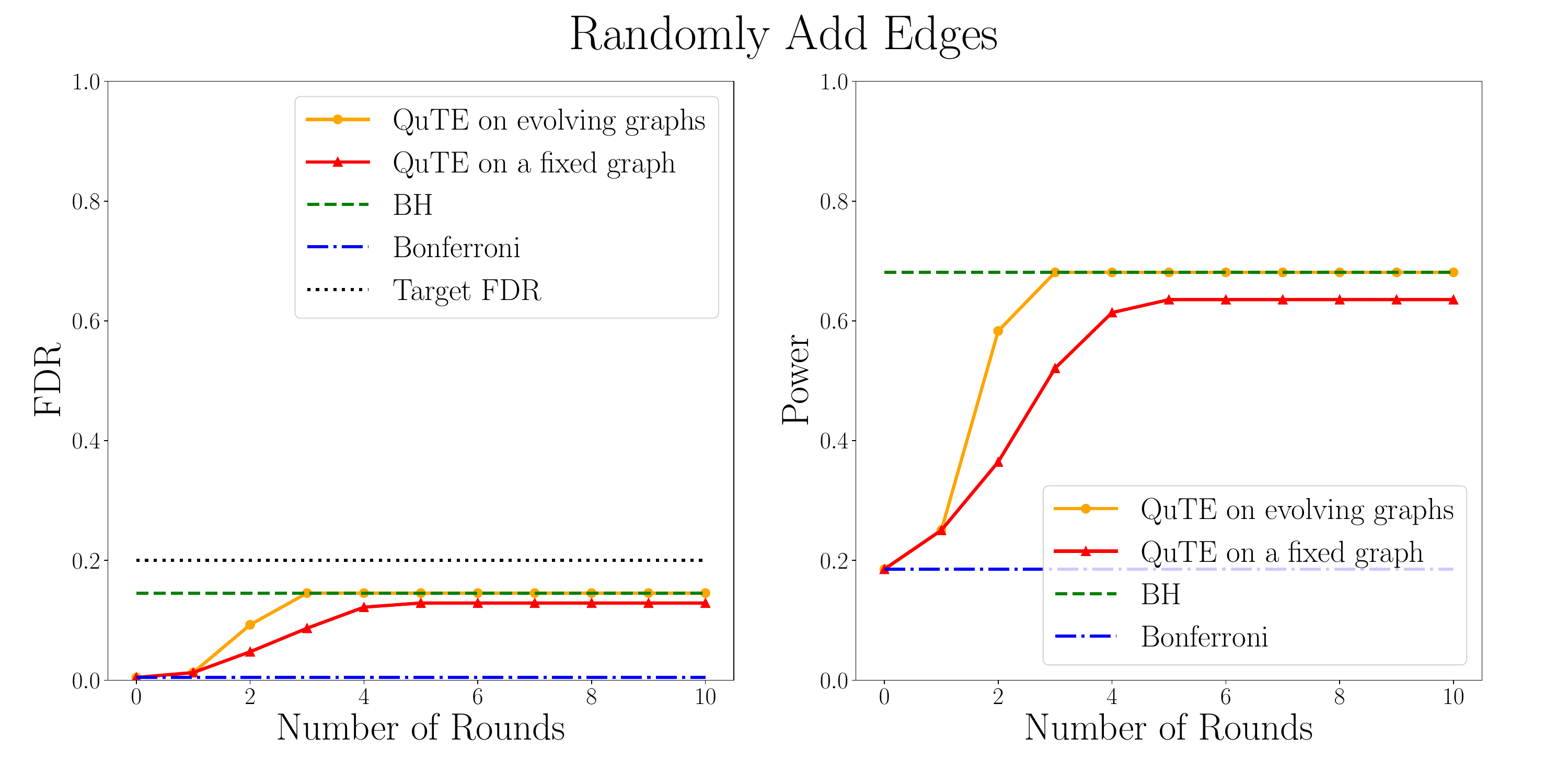}
    \includegraphics[width=0.75\linewidth]{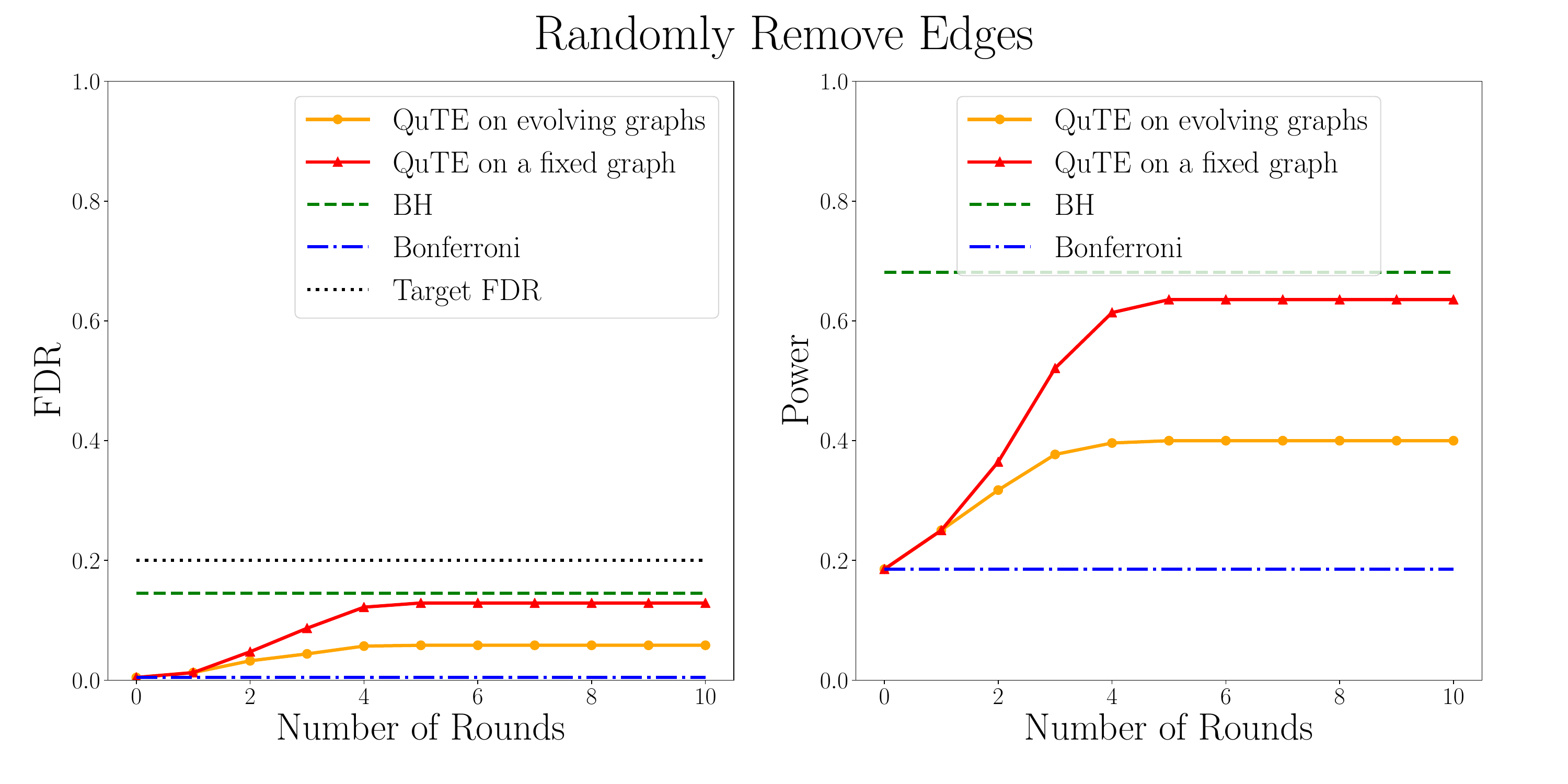}
    \includegraphics[width=0.75\linewidth]{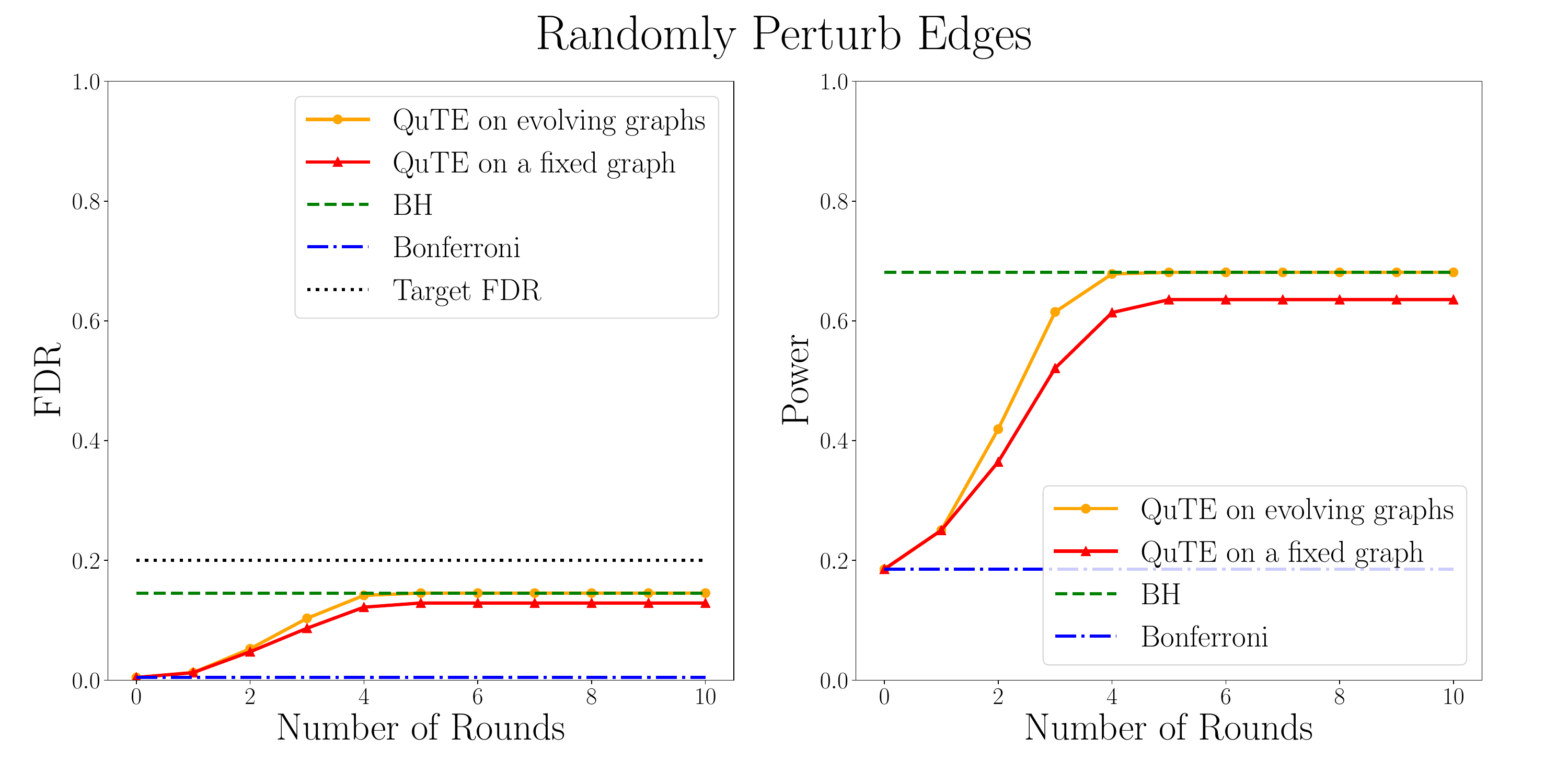}
  \caption{This figure displays the performance of QuTE when the underlying graph changes over time. The top row shows the performance of QuTE when edges are randomly added at each step, the middle row handles the case of edges being randomly deleted at each step, while the last row alternates between the two. The robustness of QuTE is apparent, as is the interpolation between the two extremes of Bonferroni (no communication) and BH (full communication).}
  \label{fig:realdata}  
\end{figure*}

\newpage
\subsection*{Acknowledgments}

This material is supported in part by the Office of
Naval Research under contract/grant number W911NF-16-1-0368.

{ \bibliography{qute} }

\bibliographystyle{abbrv}

% \newpage
\appendix

\section{Proof of \thmref{fdr}}
\label{app:prooffdr}

Let $\kh^{QuTE}$ be the total number of rejections made by the QuTE
algorithm, and let $\kh^{a}$ be the number of rejections made in the
testing phase by the agent at node $a$ in the node set of the graph
$G=(\V,\E)$. Since any rejection by the local test of any agent
results in a rejection by QuTE, it is necessarily the case that $\kh^{QuTE}
\geq \kh^{a}$ for all nodes $a \in \V$.
Next, recall that the local BH test at node $a$ is performed at level
$\alpha_a = \alpha \frac{|S_a|}{N}$. The BH test at node $a$
effectively rejects all $p$-values that are smaller than $\alpha_a
\frac{\kh^{a}}{|S_a|}$, which simplifies to $\alpha
\frac{\kh^{a}}{N}$. Now define
\begin{align*}
\small
\kh^{(a)} = \max_{s \in \text{Neighbors}(a) \cup \{a\}} \kh^s,
\end{align*}
and note that it is necessarily the case that $\kh^{QuTE} \geq
\kh^{(a)}$ for all $a$.
Recall that a $p$-value for hypothesis $j \in \{1,\dots,n_a\}$
at node $a$ is rejected either when that $p$-value is rejected by the
local test at node $a$, or when it is rejected by the local test of
one of the neighbors of $a$; hence
% \begin{align*}
$P_{a,j}$ is rejected if and only if $P_{a,j} \leq \alpha
\frac{\kh^{(a)}}{N}.$
% \end{align*}
Recalling that $\nulls_a$ is the set of true null hypotheses at
node $a$, by the definition of FDR, we have

\begin{align}
\small
  \fdr & = \EE{\dotfrac{V}{R}} 
  = \EE{\dotfrac{\sum_{a \in \V} \sum_{j \in \nulls_a} \One{P_{a,j} \leq
        \alpha \frac{\kh^{(a)}}{N}} }{ \kh }} \nonumber\\
& \leq \sum_{a \in \V} \sum_{j \in \nulls_a} \EE{\dotfrac{\One{P_{a,j}
        \leq \alpha \frac{\kh^{(a)}}{N}} }{ \kh^{(a)} }} \nonumber\\
& = \sum_{a \in \V} \sum_{j\in \nulls_a} \frac{\alpha}{N}
  \EE{\dotfrac{\One{P_{a,j} \leq \alpha \frac{\kh^{(a)}}{N}} }{ \alpha
      \frac{\kh^{(a)}}{N} }}. \label{eq:midproof}
\end{align}
Recalling that $\vec{P}_a$ is the vector of $p$-values corresponding to
hypotheses at node $a$, we denote the vector of $p$-values in the
possession of agent at node $a$ \emph{after} the querying round of QuTE as
$
\vec{P}^{(a)} := \vec{P}_a \bigcup\limits_{b \in \text{Neighbors}(a)} \vec{P}_b.
$ 
We
then define the function $f(\vec{P}^{(a)}) := \alpha
\frac{\kh^{(a)}}{N}$, and notice that the function $\vec{P}^{(a)}
\mapsto f(\vec{P}^{(a)})$ is a non-increasing function of
$\vec{P}^{(a)}$, since decreasing any of the $p$-values can only
possibly increase the number of rejections made by the neighboring
local BH tests. Hence, we may apply Lemma 1(b) from \cite{pf+}, to
conclude that
\begin{align*}
\small
\EE{\dotfrac{\One{P_{a,j} \leq \alpha \frac{\kh^{(a)}}{N}} }{ \alpha
    \frac{\kh^{(a)}}{N} }} \leq 1.
\end{align*}
Plugging this into expression \eqref{eq:midproof}, we conclude that
\begin{align*}
\small
\fdr \leq \sum_{a \in \V} \sum_{j\in \nulls_a} \frac{\alpha}{N} =
\sum_{a \in \V} \alpha \frac{|\nulls_a|}{N} = \alpha
\frac{|\nulls|}{N}.
\end{align*}

%%%%%%%%%%%%%%%%%%%%%%%%%%%%%%%%%%%%%%%%%%%%%%%%%%%%%%%%%%%%%%%%%%%%%%%%%%%%%%%%%%%%%%

\section{Proof of \propref{power}}
\label{app:proofpower}

The proof of this proposition follows from a reinterpretation of the
local BH thresholds at each node. Recall that the local BH
test at node $a$ is performed at level $\alpha^{(a)} = \alpha
\frac{|S_a|}{N}$, and hence the local BH test at node $a$ effectively
rejects any $p$-value smaller than $\alpha^{(a)} \frac{\kh^{(a)}}{|S_a|}$,
which simplifies to $\alpha \frac{\kh^{(a)}}{N}$.
Further, if one had run a fully centralized BH on all $N$ $p$-values,
we would have rejected any $p$-value smaller than $\alpha
\frac{\kh^{BH}}{N}$ for some $\kh^{BH} \geq \kh^{(a)}$. The similarity of
the rejection thresholds of centralized BH and QuTE
suggests the following: the local BH test at node $a$ on $|S_a|$
$p$-values at level $\alpha^{(a)}$ can be re-interpreted as a centralized BH test
on $N$ $p$-values at level $\alpha$ where all unknown $p$-values take the
value 1. (While this may seem pessimistic, one may note that
without further side information or prior knowledge, a local test
cannot assume anything else about the other $p$-values --- indeed, all
the other hypotheses that it knows nothing about may be true nulls,
and the distribution of the corresponding $p$-value might place a point
mass at 1, which satisfies the super-uniformity assumption required of
null $p$-values.)

The effect of adding one or more edges to a given graph, is that some
local nodes will now have more information about other $p$-values. These
$p$-values that they now possess while performing their local test might
be smaller than 1. Under the reinterpretation of QuTE, we will then be
effectively performing a local BH test at level $\alpha$ on $N$
$p$-values, more of which we have access to and fewer of which are
presumed to equal 1.  It is well known that the function $\vec{P} \mapsto \kh^{(a)}$ is
non-increasing in $\vec{P}$, the number of rejections (on the graph
with additional edges) can only possibly increase, and further the set itself (not just its size) can only
possible be enlarged. This concludes the proof.

\section{Proof of \thmref{qbh}}\label{app:proofqbh}

The proof proceeds by establishing that $\kh^Q = \kh^{BH}$. 
First, we know that $\prank_{(\kh^Q)} \leq \kh^Q$. Hence, $P_{\kh^Q} \leq \alpha \kh^Q / N$, which implies that $\kh^{BH} \geq \kh^Q$.
We also know that $P_{(\kh^{BH})} \leq \alpha \kh^{BH}/N$. Hence, $\prank_{(\kh^{BH})} \leq \kh^{BH}$, which implies that $\kh^{Q} \geq \kh^{BH}$.
Together, these establish that $\kh^Q = \kh^{BH}$. Immediately, the particular hypotheses rejected are also identical, as claimed by the theorem.

\end{document}